%% file: embc2022.tex
\documentclass[letterpaper, 10 pt, conference]{ieeeconf}
\usepackage[ieee,subfig]{definition}
\IEEEoverridecommandlockouts
\title{\LARGE \bf
Joint Embedding of Structural and Functional Brain Networks \mbox{with Graph Neural Networks for Mental Illness Diagnosis}
}

\author{
Yanqiao Zhu$^{\dagger}$, Hejie Cui$^{\ddagger}$, Lifang He$^{\S}$, Lichao Sun$^{\S}$, and Carl Yang$^{*\ddagger}$%
\thanks{*Carl Yang is the corresponding author.}%
\thanks{$^{\dagger}$Yanqiao Zhu is with the Center for Research on Intelligent Perception and Computing, Institute of Automation, Chinese Academy of Sciences, and also with School of Artificial Intelligence, University of Chinese Academy of Sciences. \texttt{\small yanqiao.zhu@cripac.ia.ac.cn}}%
\thanks{$^{\ddagger}$Hejie Cui and Carl Yang are with the Department of Computer Science, Emory University. \texttt{\small \{hejie.cui, j.carlyang\}@emory.edu}}%
\thanks{$^{\S}$Lifang He and Lichao Sun are with the Department of Computer Science and Engineering, Lehigh University. \texttt{\small \{lih319, lis221\}@lehigh.edu}}%
}

\begin{document}

\maketitle
\thispagestyle{empty}
\pagestyle{empty}

\begin{abstract}
Multimodal brain networks characterize complex connectivities among different brain regions from both structural and functional aspects and provide a new means for mental disease analysis.
Recently, Graph Neural Networks (GNNs) have become a de facto model for analyzing graph-structured data. However, how to employ GNNs to extract effective representations from brain networks in multiple modalities remains rarely explored.
Moreover, as brain networks provide no initial node features, how to design informative node attributes and leverage edge weights for GNNs to learn is left unsolved.
To this end, we develop a novel multiview GNN for multimodal brain networks. In particular, we treat each modality as a view for brain networks and employ contrastive learning for multimodal fusion.
Then, we propose a GNN model which takes advantage of the message passing scheme by propagating messages based on degree statistics and brain region connectivities.
Extensive experiments on two real-world disease datasets (HIV and Bipolar) demonstrate the effectiveness of our proposed method over state-of-the-art baselines.
\end{abstract}

\input{sections/introduction.tex}
\input{sections/method.tex}
\input{sections/experiments.tex}
\input{sections/conclusion.tex}

\section*{Acknowledgements}
This research was supported in part by the University Research Committee of Emory University, and the internal funding and GPU servers provided by the Computer Science Department of Emory University.

\bibliographystyle{IEEEbib}
\bibliography{reference}

\end{document}

%% file: sections/introduction.tex
\section{Introduction}

Mental illness is nowadays highly prevalent and has shown to be impactful for people's physical health.
With the rapid development of modern neuroimaging technology, recent years have witnessed a growing academic interest in brain network analysis, which has demonstrates its effectiveness in mental health analysis \cite{Su:2020ha}.
In neuroscience, brain networks are often represented in different modalities from structural (e.g., Diffusion Tensor Imaging, DTI) and functional aspects (e.g., functional Magnetic Resonance Imaging, fMRI) \cite{Zimmermann:2018jl}. These networked data represent complex structures of human brain connectivities. For example, in fMRI networks, edge connections represent correlations among brain regions with functional stimulations. Therefore, they are of paramount research values to understanding biologically mechanisms of brain functions.
Moreover, the existing body of research on brain network analysis suggests that different modalities of brain networks convey complementary information to each other and the fusion of multiple modalities could lead to consistent improvements for brain analysis \cite{Sun:2017hl,Ma:2017cp,Zhang:2018wo}.

Recently, Graph Neural Networks (GNNs) have emerged as a powerful tool for understanding graph-structured data in many domains \cite{Kipf:2017tc,Velickovic:2018we,Yang:2021cj}, where graph structures (i.e., adjacency matrices) and node features are embedded into a low-dimensional space for downstream machine learning.
Unlike previous shallow network embedding models that can be regarded as a certain case of matrix factorization, GNN is more powerful in terms of representation ability \cite{Xu:2019ty,Qiu:2018ez}, which makes it suitable for analyzing brain networks usually of high nonlinearity \cite{Zhang:2018wo}.

To date, there remains a paucity of studies on applying deep GNNs in analyzing multiview brain networks.
We identify two obstacles for effectively learning embeddings of brain regions in multimodal data.
Firstly, due to the multimodal nature of brain network data, different modalities encode different biomedical semantics of brain regions. How to learn effective node embeddings in such a multiview setting with GNNs remains rarely explored.
Secondly, unlike other conventional graphs such as social networks, most brain networks could be expressed in a form of weighted adjacency matrix describing connections among brain regions without initial node features. How to design informative node attributes and corresponding edge weights for GNNs to learn is left unsolved.

To address these aforementioned challenges, in this work, we develop a novel multiview framework for multimodal brain network analysis, which we refer to BrainNN for brevity.
Unlike previous work which presumes the existence of a common underlying graph structure beneath different modalities, we first treat the brain networks under different modalities as multiple views of the brain and resort to contrastive learning for jointly embedding structural and functional brain networks. Particularly, we ensure cross-view consistency by imposing a contrastive objective on node embeddings across different views. In this way, we are able to adaptively distill discriminative knowledge from each modality without the need of defining or learning a common brain network structure.

In addition, due to the lack of node and edge features in brain network analysis, we propose to derive informative attributes from the original multimodal data.
For incorporating node attributes, we take advantage of existing structural descriptors, such as degree profiles \cite{Cai:2018te} to extract features for each node.
To make full use of edge connectivity signals, contrary to directly utilizing edge weights for aggregating neighborhoods, we propose a general message passing scheme for brain networks such that we properly embed edge weights into learned node representations.
Through extensive experiments on two real-world brain disease classification datasets (HIV and Bipolar), we find that our BrainNN employing degree features and the message passing scheme with edge embeddings can consistently achieve better performance across different datasets.

 To sum up, the contribution of this work is threefold.
 \begin{itemize}
  	\item We study the applicability of GNNs on multiview brain networks in the absence of node and edge features. Specifically, we propose a general message passing GNN framework with node degree profiles as node features. 
 	\item We propose a novel multiview contrastive learning framework for brain network analysis, which adaptively extracts information from both structural and functional modalities of brain networked data.
 	\item Comprehensive experiments on two real-world brain disease classification datasets demonstrate the effectiveness of our proposed method.
 \end{itemize}

%% file: sections/method.tex

\section{The Proposed Method: BrainNN}


\subsection{Preliminaries}

\textbf{Problem definition.}
We consider the problem of multiview brain network analysis, where each brain network describes connectivities between brain regions in multiple modalities.
Suppose we are given a dataset \(\mathcal{M} = \{(\{\mathcal{G}_i^\text{s}, \mathcal{G}_i^\text{f}\}, y_i)\}_{i=1}^S\) consisting of \(S\) subjects, where \(\mathcal{G}_i^\text{s}\) and \(\mathcal{G}_i^\text{f}\) represent the structural and functional modalities of the \(i\)\textsuperscript{th} subject described in brain networks respectively and \(y_i\) is its corresponding disease label.
Each modality can be described in a weighted graph \(\mathcal{G}_i^\ast = (\mathcal{V}, \mathcal{E}_i^\ast, \bm{W}_i^\ast)\), where \(\mathcal{V} = \{v_j\}_{j = 1}^N\) is the node set of size \(N\) defined by Region Of Interests (ROIs, same across subjects), \(\mathcal{E}_i^\ast = \mathcal{V} \times \mathcal{V}\) is the edge set, and \(\bm{W}_i^\ast \in \mathbb{R}^{N \times N}\) is the weighted adjacency matrix describing interconnections between ROIs.

The purpose of multiview brain network analysis is to learn a low-dimensional representation for each subject. The learned subject embeddings can be used to facilitate disease diagnosis and treatment.

\textbf{Message passing graph neural networks.}
GNNs are widely used as a backbone for extracting features of graph-structured data. For the modeling of a node \(v_i\), a GNN involves two key components: (1) aggregating messages from its neighborhood and (2) updating its representation in the previous layer with the aggregated message. The two operations can be formulated as
\begin{align}
	\bm{a}_i^{(l)} & = \operatorname{agg}^{(l)}\left(\operatorname{msg}^{(l)}\left(\left\{\left.\bm{h}_j^{(l - 1)} \right\vert v_j \in \mathcal{N}_i \cup \{v_i\} \right\}\right)\right), \\
	\bm{h}_i^{(l)} & = \operatorname{upd}^{(l)}\left(\bm{h}_i^{(l - 1)}, \bm{a}_i^{(l)}\right).
\end{align}
For notation simplicity, we denote a GNN model containing \(L\) stacked layers as \(f^{(L)}\) thereafter.
For a graph \(\mathcal{G}\), an extra readout function \(g\) is required to obtain a graph-level embedding:
\begin{equation}
	\bm{z} = g\left(\left\{\left.\bm{h}_i \right\vert v_i \in \mathcal{V} \right\}\right),
\end{equation}
where \(\bm{h}_i = f^{(L)}(\mathcal{G}_i)\) denotes the embedding for node \(i\).
In our implementation, we resort to sum pooling given its better representation ability \cite{Xu:2019ty}.

\begin{figure}
	\centering
	\includegraphics[width=\linewidth]{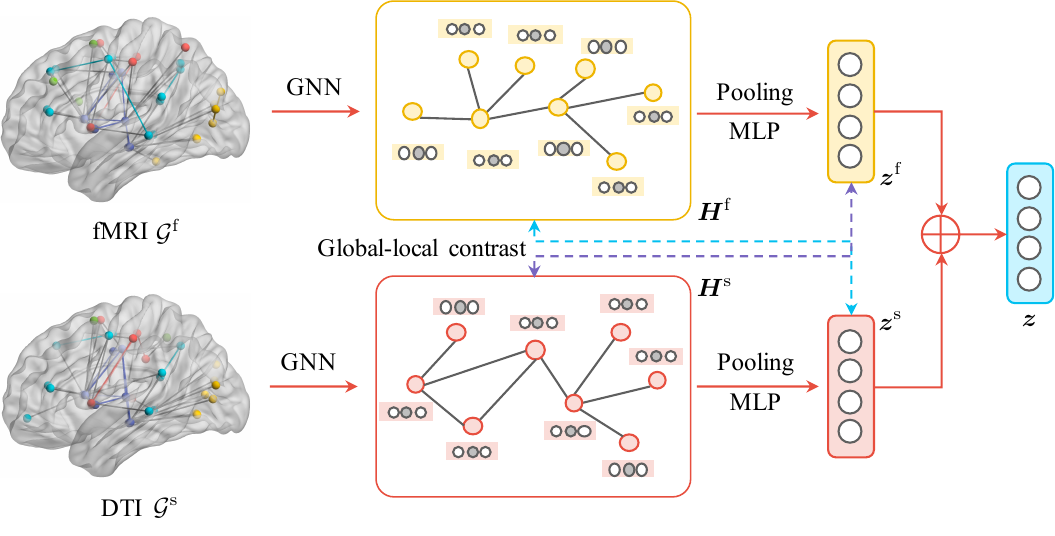}
	\caption{Our proposed BrainNN composed of two stages: graph representation learning and multimodal fusion.}
	\label{fig:model}
\end{figure}

\subsection{The Overall Framework}

Figure \ref{fig:model} summarizes the overall framework of our proposed BrainNN method. We employ contrastive learning for aggregating information from multiple modalities (\S \ref{sec:CL}), where each modality is modeled as a single view network by a specialized message passing GNN (\S \ref{sec:GNN}).

Initially, we construct node features via local statistics for each view of brain network and feed each modality graph into a GNN model; the resulting node-level embeddings are aggregated using a shared MultiLayer Perception (MLP) model to jointly learn structural and functional embeddings.
After that, we enforce cross-view consistency to adaptively integrate information from the two views by contrasting local and global representations.

\subsection{Multimodal Fusion for Structural and Functional Views}
\label{sec:CL}

We compute graph embeddings for each subject and denote the embeddings for functional and structural views as
\begin{align}
	\bm{z}_i^\text{f} & = g(f^{(L)}(\mathcal{G}_i^\text{f})), \\
	\bm{z}_i^\text{s} & = g(f^{(L)}(\mathcal{G}_i^\text{s})).
\end{align}
At the inference time, we aggregate the representations from the two views and take the average of them \(\bm{z}_i = (\bm{z}_i^\text{f} + \bm{z}_i^\text{s})/ 2\) as the graph representations for downstream tasks.

After that, inspired by recent success of graph contrastive learning \cite{Velickovic:2019tu,Zhu:2020vf,Zhu:2021vs,Zhu:2021gx}, we learn the model parameters by optimizing a contrastive objective that distinguishes node representations of one view with graph representations of the other and vice versa:
\begin{equation}
	\mathcal{J}_\text{con} = \frac{1}{2S} \sum_{\mathcal{G}_i \in \mathcal{M}} \left[ \frac{1}{N} \sum_{v_j \in \mathcal{V}} \left( I(\bm{h}_j^\text{f}; \bm{z}_i^\text{s}) + I(\bm{h}_j^\text{s}; \bm{z}_i^\text{f}) \right) \right].
	\label{eq:objective}
\end{equation}
This global-local contrasting mode encourages multi-scale consistency of graph representations \cite{Hassani:2020un,Velickovic:2018we}.
For optimization efficiency, we estimate the mutual information \(I(X; Y)\) in Eq. (\ref{eq:objective}) using Jason-Shannon Divergence (JSD) \cite{Hjelm:2019uk} described as
\begin{multline}
	I(\bm{h}_i; \bm{z}_i) = -\operatorname{sp}(-d(\bm{h}_i, \bm{z}_i)) \\
		- \frac{1}{N - 1} \sum_{v_j \in \mathcal{V} \backslash \{v_i\}} \operatorname{sp} (d(\bm{h}_i, \bm{z}_j)),
\end{multline}
where \(\operatorname{sp}(\cdot) = \log(1 + e^\cdot)\) and \(d(\cdot, \cdot)\) is a discriminator function, taking the inner product of node embeddings with a sigmoid activation.
Finally, we train the model with the unsupervised objective \(\mathcal{J}_\text{con}\) along with the supervised cross-entropy loss.

\subsection{Message Passing GNNs for Brain Networks}
\label{sec:GNN}

\textbf{Constructing node features for modality graphs.}
Non-attributed brain networks for graph classification bring challenges for applying graph neural network techniques.
We take advantages of existing local statistical measures such as degree profiles. In particular, we study Local Degree Profiles (LDP) \cite{Cai:2018te} for each brain modality, where each feature \(\bm{x}_n\) of modality graph \(\mathcal{G}_{ij}\) is computed as
\begin{equation}
	\bm{x}_n = \left[ \deg(n); \min(\mathcal{D}_n); \max(\mathcal{D}_n); \operatorname{mean}(\mathcal{D}_n); \operatorname{std}(\mathcal{D}_n) \right],
\end{equation}
where \(\mathcal{D}_n = \{\deg(m) \mid (n,m) \in \mathcal{E}_{ij}\}\) describes the degree statistics of node \(n\)'s one-hop neighborhood and \([\cdot ; \cdot]\) denotes concatenation.
Such computation can be done in \(O(\mathcal{E})\) time, which is computational friendly.

\textbf{Handling edge weights.}
After obtaining node features for each modality graph, we feed it into a message passing GNNs, with parameters shared across all modalities.
For notation simplicity, we focus on one modality graph in this section and thus omit the subscript referring to specific graphs.
Since the brain region connectivity is expressed in edge weights, we first construct a message vector \(\bm{m}_{ij} \in \mathbb{R}^{D}\) composed of node embeddings of a node \(i\), its neighborhood \(j\), and edge weight \(w_{ij}\):
\begin{equation}
	\bm{m}_{ij}^{(l)} = t_{\bm\Theta} \left( \left[ \bm{h}_i^{(l)};\, \bm{h}_j^{(l)};\, w_{ij} \right] \right),
\end{equation}
where \(t_{\bm\Theta}\) denotes a MLP layer parameterized by \(\bm\Theta\), and \(l\) is the index of the current GNN layer.

Then, we aggregate messages from all neighborhoods followed by a non-linear transformation; the node-wise propagation rule can be written as:
\begin{equation}
	\bm{h}_i^{(l)} = \sigma \left( \sum_{j \in \mathcal{N}_i \cup \{i\}} \bm{m}_{ij}^{(l - 1)} \right),
\end{equation}
where \(\sigma\) is a non-linear activation function such as \(\operatorname{ReLU}(\cdot) = \max(0, \cdot)\).

Finally, we summarize all node embeddings using sum pooling and employ another MLP parameterized by \(\bm\Phi\) with residual connections \cite{He:2016ib} to compute graph-level embeddings \(\bm{z} \in \mathbb{R}^{D}\):
\begin{align}
	\bm{z}' & = \sum_{i \in \mathcal{V}} \bm{h}_{i}^{(k)}, \\
	\bm{z} & = t_{\bm\Phi} ( \bm{z}' ) + \bm{z}'.
\end{align}
The final representation vector \(\bm{z}\) extracts essential information of one subject and thus could be used for disease diagnosis.

%% file: sections/experiments.tex
\section{Experiments}

\subsection{Datasets and Data Preprocessing}

In the experiments, we use two datasets collected by The University of Chicago to evaluate the effectiveness of our method:
\begin{itemize}
	\item \textbf{Human Immunodeficiency Virus Infection (HIV)} contains 35 patients (positive) and 35 seronegative controls (negative). Each modality graph is with 90 nodes and edge weights are calculated as correlations between brain regions.
	The HIV dataset is constructed following Cao et al. \cite{Cao:2015kf}.
	For the fMRI data, we use the DPARSF toolbox\footnote{\url{http://rfmri.org/DPARSF}} to conduct realignment, time correction, normalization, and signal smoothing.
	For the DTI data, we make use of the FSL toolbox\footnote{\url{http://fsl.fmrib.ox.ac.uk/fsl/fslwiki}} for preprocessing the original data, involving distortion correction, noise filtering, and repetitive sampling.
	
	\item \textbf{Bipolar Disorder (BP)} consists of 52 bipolar subjects in euthymia and 45 healthy controls with matched age and gender. It stimulates 82 brain regions, according to Freesurfer-generated cortical/subcortical gray matter regions.
	We use the CONN toolbox\footnote{\url{http://www.nitrc.org/projects/conn}} to construct the brain network. Specifically, we realigned and co-registered the raw images, after which we perform normalization and smoothing.
\end{itemize}

\subsection{Experimental Protocols}

\textbf{Metrics.}
In experiments, we report the classification performance in terms of Accuracy and Area Under the ROC Curve (AUC), which are widely used for disease identification. Larger values indicate better performance.

\textbf{Baselines.}
We include a broad range of baseline methods for comprehensive evaluation.
For shallow embeddings methods, our experiments include
\begin{itemize}
	\item M2E \cite{Liu:2018ty} leverages tensor-based multimodal fusion to obtain embeddings. We apply M2E on all subjects and leverage a fully connected network for classification.
	\item MIC \cite{Shao:2015ek} conducts kernel decomposition to extract feature representations for subjects. Similar to M2E, we apply a fully connected network on the resulting embedding for classification.
	\item MPCA \cite{Lu:2008cw} is a general approach to extract features from tensor objects. In our setting, we concatenate all features to form a 4D tensor and then apply MPCA to obtain the embeddings for each subject across modalities and individuals.
	\item MK-SVM \cite{Dyrba:2015ci} leverages the SVM classifier on multiple modalities of the brain imaging data.
\end{itemize}

We also consider the following up-to-date deep learning models:
\begin{itemize}
	\item 3D-CNN \cite{Gupta:2013tp} leverages sparse autoencoder with convolutional networks on neural images. We concatenate fMRI and DTI modalities into a 3D tensor and apply 3D-CNN to enable end-to-end training.
	\item GCN \cite{Kipf:2017tc} generalizes convolutional operations into graph domains. We first construct a feature matrix, where each row corresponds to the vectorized representation of 3D multimodal data. Then, we regard each vectorized graph as a node and directly apply GCN.
	\item GAT \cite{Velickovic:2019tu} introduces attention mechanisms to GCN. Similarly, we apply the GAT model on the graphs.
	\item DiffPool \cite{Ying:2018vl} is a hierarchical GCN model with differentiable pooling. Similarly, we use it on the graphs.
	\item MVGCN \cite{Zhang:2018wo} leverages multiple views for GCN. We obtain the shared feature space by averaging all brain network across modalities and subjects.
\end{itemize}
Among them, 3D-CNN and MVGCN are designed for multiview learning and we perform classification in an end-to-end manner. For the other three baselines, we apply it with each single modality and report the best performance.

\subsection{Implementation Details}
For the two datasets, we closely follow the experimental setting of prior work \cite{Zhang:2018wo}. Specifically, we use 80\% of data for training, 10\% for validation, and the remaining 10\% for test.
In our BrainNN model, we apply grid search to determine the optimal hyperparameters.
In particular, we empirically select the embedding dimension among \{20, 40, 60, 80, 100\} and the number of GNN layers among \{1, 2, 3\}.

\subsection{Results and Analysis}

\begin{table}
	\small
	\centering
	\caption{Comparison of different models on HIV and BP datasets. The highest performance is highlighted in boldface and passed significant tests with \(p \leq 0.05\).}
	\begin{tabular}{ccccc}
	\toprule
	\multirow{2.5}{*}{Method} & \multicolumn{2}{c}{HIV} & \multicolumn{2}{c}{BP} \\
	\cmidrule(lr){2-3} \cmidrule(lr){4-5}
		& Accuracy & AUC   & Accuracy & AUC \\
	\midrule
	M2E   & 50.61 & 51.53 & 57.78 & 53.63 \\
	MIC   & 55.63 & 56.61 & 51.21 & 50.12 \\
	MPCA  & 67.24 & 66.92 & 56.92 & 56.86 \\
	MK-SVM & 65.71 & 68.89& 60.12 & 56.78 \\
	\midrule
	3D-CNN & 74.31 & 73.53 & 63.33 & 61.62 \\
	GAT   & 68.58 & 67.31 & 61.31 & 59.93 \\
	GCN   & 70.16 & 69.94 & 64.44 & 64.24 \\
	DiffPool & 71.42 & 71.08 & 62.22 & 62.54 \\
	MVGCN & 74.29 & 73.75 & 62.22 & 62.64 \\
	\midrule
	\rowcolor{lightgray!20} BrainNN & \textbf{77.14} & \textbf{79.79} & \textbf{73.64} & \textbf{67.54}\\
	\bottomrule
	\end{tabular}
	\label{tab:performance}
\end{table}

The overall performance is presented in Table \ref{tab:performance}. It is apparently seen that our proposed method achieves the best in two datasets. To be specific, BrainNN outperforms previous graph-based baselines MVGCN and DiffPool by large margins, up to 11\% absolute improvements, which demonstrates the effectiveness of multiview fusion and the applicability of message passing GNN on brain networks.
Compared to traditional tensor-based methods, such as M2E and MPCA, our work achieves significantly better results, which verifies the necessity of incorporating deep GNN for learning informative graph representations.
The rationale of jointly embedding structural and functional brain networks can be further supported by the superiority of our work compared with deep models designed for single view such as GCN and GAT.
Moreover, we notice that 3D-CNN obtains promising results, owing to the ability of modeling locality of input features. However, it computes graph representations with mere pooling layers, which fails to model the interaction among different views. Our work, on the contrary, leverages contrastive learning to model interaction of structural and functional views of brain networks, leading to discriminative representations in an adaptive manner.

\begin{table}
	\small
	\centering
	\caption{Performance of ablated models on HIV and BP datasets. Results passed significant tests with \(p \leq 0.05\).}
	\begin{tabular}{ccccc}
	\toprule
	\multirow{2.5}{*}{Method} & \multicolumn{2}{c}{HIV} & \multicolumn{2}{c}{BP} \\
	\cmidrule(lr){2-3} \cmidrule(lr){4-5}
		& Accuracy & AUC   & Accuracy & AUC \\
	\midrule
	V-GCN & 70.00 & 75.83 & 67.14 & 61.17\\
	CONCAT & 66.36 & 72.39 & 67.27 & 61.13 \\
	\rowcolor{lightgray!20} BrainNN & \textbf{77.14} & \textbf{79.79} & \textbf{73.64} & \textbf{67.54}\\
	\bottomrule
	\end{tabular}
	\label{tab:ablation}
\end{table}

\subsection{Ablation Studies}

We further verify the effectiveness of two key components in our model: message passing GNN and multiview fusion based on contrastive learning.
Firstly, to examine the impact of the message passing GNN, we compare it with a vanilla Graph Convolutional Network model (\textbf{V-GCN}). The V-GCN directly treats the correlation weights as the adjacency matrix and performs weighted spectral convolution, which does not properly handle edge weights.
Secondly, to validate the effectiveness of our multimodal fusion scheme, we further compare it against a simple baseline model named \textbf{CONCAT} that concatenates embeddings of all two views without multiview fusion.
For fair comparison, all other experimental configurations are kept the same as in previous section. The results are summarized in Table \ref{tab:ablation}. We observe that on the two datasets, our BrainNN method outperforms its two downgraded versions.
It is worthy noting that the V-GCN still achieves close-to-optimal performance, demonstrating the effectiveness of applying deep GNNs with multiview interaction.

\subsection{Visualization}

\begin{figure}
	\centering
	\subfloat[fMRI]{
		\includegraphics[width=\linewidth]{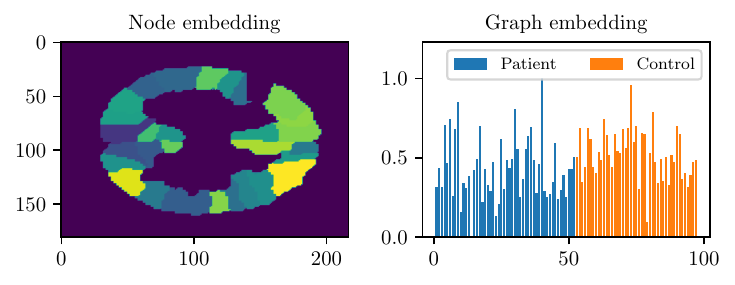}
	}\\
	\subfloat[DTI]{
		\includegraphics[width=\linewidth]{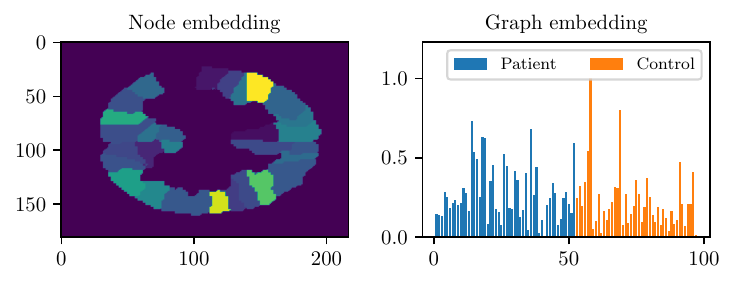}
	}
	\caption{Visualization of embedded features from fMRI and DTI on the BP dataset. Left: learned node embeddings \(\bm{H}\) with color denoting the intensity of brain activities. Right: learned graph embeddings \(\bm{Z}\), with the index of subjects across patients and healthy controls shown on the \(x\)-axis and the corresponding embedding values on the \(y\)-axis.}
	\label{fig:visualization}
\end{figure}

To qualitatively examine the effectiveness of our method, we visualize the embeddings learned by our model from clinical perspective.
In Figure \ref{fig:visualization}, we plot the learned node embeddings in the left panel, where the coordinate system represents neuroanatomy and color denotes the intensity of brain activities.
It is seen from the figure that node embeddings from structural and functional perspectives demonstrate wide discrepancies to each other.

We also present the visualization for graph embeddings representing the factor strengths for both patients and health controls in the right panel.
We observe a relatively positive correlation in the embeddings of the control group, while the patients have a relatively negative correlation with these factors. It is thus evidently necessary to properly model and combine both views to deliver ideal clinical diagnoses.

%% file: sections/conclusion.tex
\section{Conclusion}
In this work, we have proposed a novel BrainNN framework that jointly embeds multimodal brain networks with GNNs for mental illness diagnosis.
Extensive experiments on two real-world datasets demonstrate the effectiveness of our proposed method.
The study of applying GNNs for brain networks remains widely open with many challenges left to solve. For example, the current databases of small scales greatly confine the training of deep GNN models, for which we plan to investigate transfer learning and pre-training techniques \cite{Hu:2020uz,Hu:2020vh} in the near future.